# A Post-Quantum Key Agreement Protocol Based on a Modified Matrix-Power Function over a Rectangular Integer Matrices Semiring

**Juan Pedro Hecht [1], Hugo Daniel Scolnik [2],***

[1] Master's degree in Information Security, Economic Science School, Exact and Natural Sciences School and Engineering School, University of Buenos Aires, Av. Cordoba 2122, C1120AAP, Ciudad Autónoma de Buenos Aires, Argentina; phecht@dc.uba.ar

[2] Master's degree in Information Security, Economic Science School, Exact and Natural Sciences School and Engineering School, University of Buenos Aires, Av. Cordoba 2122, C1120AAP, Ciudad Autónoma de Buenos Aires, Argentina; hugo@dc.uba.ar

* Correspondence: hscolnik@gmail.com ; Tel.: +54-911-4970-6665

**Abstract:** We present an improved post-quantum version of Sakalauskas matrix power function key agreement protocol, using rectangular matrices instead of the original square ones. Sakalauskas matrix power function is an efficient and secure way to generate a shared secret key, and using rectangular matrices provides additional flexibility and security. This method reduces the computational complexity by allowing smaller random integer matrices while maintaining a high level of security. We don't rely on matrices with special formatting to achieve commutativity; instead, we use full random values on those structures, increasing their entropy. Another advantage of using rectangular matrices over key agreement protocols is that they offer better protection against various linearization attacks.

## 1. Introduction

Post-quantum cryptography has become an important area of research. It aims to develop cryptographic algorithms that are secure against attacks by quantum computers [1]. One of the most important applications of post-quantum cryptography is key agreement protocols [2], which allow two parties to establish a shared secret key over an insecure channel. In this paper, we explore the use of rectangular matrices instead of traditional square ones in a post-quantum key agreement protocol using the Sakalauskas matrix power function [3-8].

## 2. Paper organization

First, we define the rectangular matrix power function (RMPF), a generalization of the matrix power function (MPF) introduced by Sakalauskas [3-8], and describe its properties. Second, the key agreement protocol (KAP) based on the RMPF is presented. Then, a simplified numerical example of the protocol is given, followed by security considerations, and ending with a discussion of the advantages of using rectangular matrices in the given protocol.

## 3. Definitions and properties of the Rectangular Matrix Power Function (RMPF)

We use here the same notation used in the work of Sakalauskas [3-8].

**Definition 1.** *Equidimensional (m,n) rectangular matrices of integers (specifically p-prime $\mathbb{Z}p$ field elements) form an RM set, a ring structure with p-modular sums and p-modular Hadamard products (Modular operations keep numbers under constant format).*

**Definition 2.** *Matrix elements of RM-set n-powers are formed with p-modular n-powers of each element of the base matrix. Therefore, the product of a x-power of W element by a y-power of the same element commute (Wx.Wy= Wy.Wx) since the integer exponents product x, y commute. From now on, this paper deals only with RM-sets, whenever rectangular matrices are invoked.*

**Definition 3.** *Given any three matrices (X,W,C) of the same (m, n) RM-set, the left-sided rectangular matrix power function (RMPF) exponential action of X over W, is defined as the matrix C= $\{c_{ij}\}$:*

$$X \triangleright W \equiv {}^XW = C \quad \text{where} \quad c_{ij} = \prod_{k=1}^{rank[X]} w_{kj}{}^{x_{ik}} \tag{1}$$

**Definition 4.** *Given any three matrices (W, Y, D) of the same (m, n) RM-set, the right RMPF exponential action of Y over W, is equal to D; $D = \{d_{ij}\}$:*

$$W \triangleleft Y \equiv W^Y = D \quad \text{where} \quad d_{ij} = \prod_{l=1}^{rank[Y]} w_{il}{}^{y_{lj}} \tag{2}$$

**Definition 5.** *Given any four matrices (X, W, Y, Q) of the same (m, n) RM-set, the double-sided RPMF exponential action of the matrix W with the left-sided X - matrix action and the right-sided Y matrix action is defined as Q; $Q = \{q_{ij}\}$:*

$$X \triangleright W \triangleleft Y \equiv {}^XW^Y = Q \quad \text{where} \quad q_{ij} = \prod_{k=1}^{rank(X)} \prod_{l=1}^{rank[Y]} w_{kl}{}^{x_{ik} \cdot y_{lj}} \tag{3}$$

**Lemma 1**. *The RPMF is unilaterally associative, as Sakalauskas proved [7], if the following identities hold:*

$$^Y({}^XW) = {}^{(YX)}W = {}^{YX}W \; ; \; (W^X)^Y = W^{(XY)} = W^{XY} \tag{4}$$

*and two-sided associative if:*

$$({}^XW)^Y = {}^X(W^Y) = {}^XW^Y \tag{5}$$

*and RMPF is defined as associative if both conditions hold.*

**Lemma 2.** *(m, n) RM-sets, obey the associative properties of RMPF. This is a special case of Sakalauskas proof [14 and others], since the square (m, m) matrices are replaced by the particular case of (m,n) rectangular ones.*

□

**Lemma 3.** *If (X, Y, U, V, W) are (m, n) RM-set matrices acting as one-sided (left or right) RMPF actions over another W and (X, U), (Y, V) pairs respectively outer (ordinary) products, then both satisfy the commutative conditions:*

$$X^T.U = U^T.X \; ; \; Y^T.V = V^T.Y \tag{6}$$

*and RMPF over RM sets are associative (eq 4,5), then:*

$$^U({}^XW^Y)^V = {}^{UX}W^{YV} = {}^{XU}W^{VY} = {}^X({}^UW^V)^Y \tag{7}$$

**Proof.** *If pairwise outer products commute, the elements of their square product matrix exponents could be interchanged (see Definition 2. properties applied to (1), (2), (3) equations). Therefore, equation (7) holds.* □

**Lemma 4**. *If $(\lambda 1, \lambda 2) \in \mathbb{Z}^2$ and $(X, U, W)$ are members of the same RM-set, then the scalar products $\lambda 1.W = X$ and $\lambda 2.W = U$ are matrices satisfying condition (6).*

**Proof.** *Given an RM-set matrix $W = \{w_{ij}\}$, then $\lambda W = \{\lambda w_{ij}\}$ and $oW = \{ow_{ij}\}$, as $(\lambda, o) \in \mathbb{Z}2$ then $\lambda w_{ij}.ow_{ij} = ow_{ij}.\lambda w_{ij} = (o.\lambda)w_{ij}^2 = (\lambda.o)w_{ij}^2$ and therefore, condition (6) holds.* □

## 4. Schematic Key Agreement Protocol (KAP) using the proposed RMPF action

| RMPF PROTOCOL | |
|---|---|
| **ALICE** | **BOB** |
| **Define shared public parameters:**<br>p: random prime (64 bits)<br>rows, cols: dimensions of rectangular matrices (rows > cols)<br>Base, X, Y: random rectangular integer matrices (values in $\mathbb{Z}_p$) | |
| **Private values:**<br>lambdaA, omegaA:<br>random integers in $\mathbb{Z}_p$<br>A1=lambdaA.X<br>B1=omegaA.Y | **Private values:**<br>lambdaB, omegaB:<br>random integers in $\mathbb{Z}_p$<br>A2=lambdaB.X<br>B2=omegaB.Y |
| **Public value:**<br>$TA$ = matrix wherein<br>$TA_{ij} = \prod_{l=1}^{rank[A1]} \prod_{k=1}^{rank[B1]} Base^{A1_{ik} \cdot B1_{lj}} \pmod{p}$ $\longrightarrow$ $TA$ | |
| | **Public value:**<br>$TB$ $\longleftarrow$ $TB$ = matrix wherein<br>$TB_{ij} = \prod_{l=1}^{rank[A2]} \prod_{k=1}^{rank[B2]} Base^{A2_{ik} \cdot B2_{lj}} \pmod{p}$ |
| **Key:**<br>$KeyA$ = matrix wherein<br>$keyA_{ij} = \prod_{l=1}^{rank[A1]} \prod_{k=1}^{rank[B1]} TB^{A1_{ik} \cdot B1_{lj}} \pmod{p}$ | **Key:**<br>$KeyB$ = matrix wherein<br>$keyB_{ij} = \prod_{l=1}^{rank[A2]} \prod_{k=1}^{rank[B2]} TA^{A2_{ik} \cdot B2_{lj}} \pmod{p}$ |
| **KeyA = KeyB** | |

**Figure 1.** This scheme shows the Key Agreement Protocol here proposed which is based on the RMPF.

**Lemma 5.** $keyA = keyB$

**Proof.** Considering (6) and (7), $KeyA = A1 \triangleright TB \triangleleft B1 = A1 \triangleright (A2 \triangleright Base \triangleleft B2) \triangleleft B1 = A2 \triangleright (A1 \triangleright Base \triangleleft B1) \triangleleft B2 = A2 \triangleright TA \triangleleft B2 = KeyB$. □

## 5. Detailed Key Agreement Protocol (KAP) using the proposed RMPF action

*5.1. Setup*

Both parties (Alice and Bob) agree on:
1. A random prime p (minimum 64 bit).
2. RM set dimensions (m, n), where m>n.
3. Three RM random matrices Base, X, Y with values in $\mathbb{Z}_p$ are shared between them.

*5.2. Alice's private values*

4. lambdaA, omegaA: random numbers in $\mathbb{Z}_p$
5. A1=lambdaA.X ; B1=omegaA.Y

*5.3 Alice's public token*

6. Generate the TA matrix; {TAij }
7. $TA_{ij} = \prod_{l=1}^{rank[A1]} \prod_{k=1}^{rank[B1]} Base^{A1_{ik} \cdot B1_{lj}} \pmod{p}$

8. TA is sent to Bob.

*5.4 Bob's private values*

9. lambdaB, omegaB: random numbers in $\mathbb{Z}_p$
10. A2=lambdaB.X ; B2=omegaB.Y

*5.5 Bob's public token*

11. Generate the TB matrix; {TBij }
12. $TB_{ij} = \prod_{l=1}^{rank[A2]} \prod_{k=1}^{rank[B2]} Base^{A2_{ik} \cdot B2_{lj}} \pmod{p}$
13. TB is sent to Alice.

*5.5 Shared key*

14. $keyA_{ij} = \prod_{l=1}^{rank[A1]} \prod_{k=1}^{rank[B1]} TB^{A1_{ik} \cdot B1_{lj}} \pmod{p}$
15. $keyB_{ij} = \prod_{l=1}^{rank[A2]} \prod_{k=1}^{rank[B2]} TA^{A2_{ik} \cdot B2_{lj}} \pmod{p}$

## 6. A toy numerical example (KAP)

*6.1. Small KAP full description*

Defining prime p = 104729. it follows:

```
rows = 5
cols = 3
        ⎛ 51141  16202  66646 ⎞
        ⎜  4601  73510   9641 ⎟
Base =  ⎜ 41977  29822  28262 ⎟
        ⎜ 61281  20522  40337 ⎟
        ⎝ 25689  35123  17039 ⎠

    ⎛ 27536  23259   3230 ⎞
    ⎜ 97577  61064  52197 ⎟
X = ⎜ 61356  19870  66794 ⎟
    ⎜ 93047  74112  73769 ⎟
    ⎝ 88730  84531  46584 ⎠

    ⎛  7991  99112  88031 ⎞
    ⎜ 62951  45825  26429 ⎟
Y = ⎜ 53671  81823  10939 ⎟
    ⎜ 92791  39779 100242 ⎟
    ⎝ 67646  52695  65391 ⎠
```

**Figure 2.** Setup of public values for Alice and Bob.

```
lambdaA = 35413
     ⎛  975132368   823670967   114383990 ⎞
     ⎜ 3455494301  2162459432  1848452361 ⎟
A1 = ⎜ 2172800028   703656310  2365375922 ⎟
     ⎜ 3295073411  2624528256  2612381597 ⎟
     ⎝ 3142195490  2993496303  1649679192 ⎠

omegaA = 22911
     ⎛  183081801  2270755032  2016878241 ⎞
     ⎜ 1442270361  1049896575   605514819 ⎟
B1 = ⎜ 1229656281  1874646753   250623429 ⎟
     ⎜ 2125934601   911376669  2296644462 ⎟
     ⎝ 1549837506  1207295145  1498173201 ⎠
```

**Figure 3.** Alice's private values.

```
         ⎛ 90444  78140  22111 ⎞
         ⎜ 91141  86834  31963 ⎟
TokenA = ⎜ 22517  82376  27232 ⎟
         ⎜ 76737  17315  37169 ⎟
         ⎝ 95799  99846  20180 ⎠
```

**Figure 4.** Alice's public token.

```
lambdaB = 77591
```

$$A2 = \begin{pmatrix} 2\,136\,545\,776 & 1\,804\,689\,069 & 250\,618\,930 \\ 7\,571\,097\,007 & 4\,738\,016\,824 & 4\,050\,017\,427 \\ 4\,760\,673\,396 & 1\,541\,733\,170 & 5\,182\,613\,254 \\ 7\,219\,609\,777 & 5\,750\,424\,192 & 5\,723\,810\,479 \\ 6\,884\,649\,430 & 6\,558\,844\,821 & 3\,614\,499\,144 \end{pmatrix}$$

```
omegaB = 9608
```

$$B2 = \begin{pmatrix} 76\,777\,528 & 952\,268\,096 & 845\,801\,848 \\ 604\,833\,208 & 440\,286\,600 & 253\,929\,832 \\ 515\,670\,968 & 786\,155\,384 & 105\,101\,912 \\ 891\,535\,928 & 382\,196\,632 & 963\,125\,136 \\ 649\,942\,768 & 506\,293\,560 & 628\,276\,728 \end{pmatrix}$$

**Figure 5.** Bob's private values.

$$TokenB = \begin{pmatrix} 25\,880 & 18\,100 & 3262 \\ 66\,621 & 6366 & 37\,099 \\ 77\,233 & 4706 & 92\,229 \\ 41\,946 & 98\,748 & 61\,670 \\ 61\,540 & 92\,962 & 89\,447 \end{pmatrix}$$

**Figure 6.** Bob's public token.

$$KeyA = \begin{pmatrix} 76\,099 & 14\,814 & 8343 \\ 58\,724 & 39\,308 & 74\,495 \\ 26\,031 & 18\,945 & 38\,075 \\ 90\,635 & 51\,524 & 65\,266 \\ 23\,296 & 83\,580 & 22\,846 \end{pmatrix}$$

$$KeyB = \begin{pmatrix} 76\,099 & 14\,814 & 8343 \\ 58\,724 & 39\,308 & 74\,495 \\ 26\,031 & 18\,945 & 38\,075 \\ 90\,635 & 51\,524 & 65\,266 \\ 23\,296 & 83\,580 & 22\,846 \end{pmatrix}$$

**Figure 7.** Shared keys.

*6.2. Real life parameters and security.*

Sakalauskas [8] suggests that the success rate of a brute force attack decreases exponentially as the matrix order increases. In our context, this is irrelevant since the (X, Y) matrices are public and the security relies upon lambda and omega secret integers. Therefore, much attention must be paid to the pseudo-random number generator, since the security of the protocol depends sensibly on it, given the linear relationship between the public parameters (X, Y) and the private values (A, B).

Since an attacker does not know (lambda, omega, A, B), a natural attack would be the systematic exploration of the space of the random constants (lambda, omega) which depend directly on the cardinal of the set $\mathbb{Z}_p$ and in consequence the security against this attack is proportional to $p^2$. We recommend using $p \sim 2^{64}$ as a minimum value. Thus, two random integers in $\mathbb{Z}_p$ represent a 128-bit brute-force search.

Consequently, overall security relies on the NP-hard nonlinear MPF [2] if the linear step becomes practically invulnerable.

## 7. Discussion

*7.1. Background.*

The advent of quantum computing poses a significant threat to the security of current cryptographic protocols [1]. Therefore, post-quantum cryptography has become an active area of research to develop cryptographic algorithms that can withstand attacks from quantum computers. Here we rely on the key agreement protocol developed by Sakalauskas, the Matrix Power Function (MPF) [3-8]. This is an NP-hard one-way (trapdoor) function [2] that has proven over time to be efficient and secure for generating shared secret keys. No useful attack against the use of the enhanced MPF {6,8] has appeared in the years since it was first published.

*7.2. Our contribution*

Here we present:
- A variant using rectangular matrices instead of the original square matrices of the Sakalauskas MPF. Using rectangular matrices can provide additional flexibility and security, as the added singularity blocks algebraic linearization or Gröbner basis attacks [8-14]. Further research in this area is advisable to explore the full potential of rectangular matrices in post-quantum key agreement protocols [15, 16].

- Replacing standard algebraic matrix products with Hadamard products, an unavoidable change to adapt products without recourse to transposed matrices. This approach does not simplify the attacks since the nonlinearity is assured by the intrinsic MPF mechanism.
- Using p-modular operations to deal with algebraic attacks (congruence as opposed to equality) and to keep the numerical format well dimensioned.
- Increasing the entropy of key search spaces, replacing the use of circulant matrices or restricted algebraic groups [8] to achieve commutativity with unstructured random integers.

To compute the complexity order of our solution, further research is needed in this area.

**Supplementary Materials:** A Mathematica 11 notebook with all functions used in our KAP could be distributed upon request to phecht@dc.uba.ar

**Author Contributions:** Conceptualization and methodology: both authors contributed equally; software: Hecht.; validation, formal analysis, investigation, resources, data curation, writing, visualization, review, supervision, project administration: both authors contributed equally. All authors have read and agreed to the published version of the manuscript.

**Funding:** This research received no external funding.

**Conflicts of Interest:** The authors declare no conflicts of interest.

**References**

1. Shor P.W., Polynomial-time algorithms for prime factorization and discrete logarithms on a quantum computer. SIAM Rev. 1999, 41, 303–332.
2. Menezes A. et al, Handbook of applied cryptography, 1997, The CRC Press series on discrete mathematics and its applications
3. Sakalauskas E., Enhanced matrix power function for cryptographic primitive construction. Symmetry 2018, 10, 43.
4. Sakalauskas E. et al, Key agreement protocol (KAP) based on matrix power function. In Advanced Studies in Software and Knowledge Engineering; Information Science and Computing; 2008; pp. 92–96.
5. Sakalauskas, E.; Luksys, K. Matrix power function and its application to block cipher s-box construction, Int. J. Inn. Comp. Inf. Contr. 2012, 8, 2655–2664.
6. Sakalauskas E. and Mihalkovich A., Improved Asymmetric Cipher Based on Matrix Power Function Resistant to Linear Algebra Attack. Informatica 2017, 28, 517–524.
7. Sakalauskas E. and Mihalkovich A., MPF Problem over Modified Medial Semigroup Is NP-Complete. Symmetry 2018, 10, 571.
8. Sakalauskas E. and Mihalkovich A., A New asymmetric cipher of non-commuting cryptography class based on enhanced MPF, IET Information Security , 2020, Volume 14, Issue 4 , 410-418
9. Liu J. et al, A linear algebra attack on the non-commuting cryptography class based on matrix power function. In International Conference on Information Security and Cryptology; Springer: Cham, Switzerland, 2016; pp. 343–354. matrix power function. Informatica 2014, 25, 283–298.
10. Roman'kov V., Cryptanalysis of a combinatorial public key cryptosystem. Groups Complexity Cryptology, 9(2), 125-135, 2017.
11. Roman'kov V., A nonlinear decomposition attack. Groups Complexity Cryptology, 8(2), 197-207, 2016.
12. Myasnikov A. and Roman'kov V., A linear decomposition attack. Groups Complexity Cryptology,2015, 7(1), 81-94
13. Myasnikov A. et al, Non-commutative Cryptography and Complexity of Group-theoretic Problems, Mathematical Surveys and Monographs, 2011, AMS Volume 177
14. Ben-Zvi A. et al, Cryptanalysis via algebraic spans. In Advances in Cryptology–CRYPTO 2018: 38th Annual International Cryptology Conference, Santa Barbara, CA, USA, August 19–23, 2018, Proceedings, Part I 38 (pp. 255-274). Springer International Publishing
15. Scolnik H.D. and Hecht J.P., A New Post-Quantum Key Agreement Protocol and Derived Cryptosystem Based on Rectangular Matrices. 2022, Cryptology ePrint Archive, https://eprint.iacr.org/2022/1370
16. Scolnik H.D. and Hecht J.P., Post-Quantum Key Agreement Protocol based on Non-Square Integer Matrices, 2023, https://arxiv.org/abs/2301.01586